\documentclass[a4]{llncs}

\usepackage{booktabs}   
 
 \usepackage[pdftex]{graphicx}
 \usepackage{subfigure}

\usepackage{amsmath}
\usepackage{amsfonts}
\usepackage{amssymb}

\usepackage{balance}
\usepackage{nicefrac}

\usepackage[mathscr]{euscript}

\usepackage{color}

\begin{document}

\title{Utility-Preserving Privacy Mechanisms for Counting Queries}         
 
\author{\; Natasha Fernandes\;\;\;\; \qquad\qquad  Kacem Lefki  \qquad\qquad Catuscia Palamidessi\quad\\[-2mm]}
\vspace{-2mm}
\institute{ \quad\qquad INRIA \quad\qquad\qquad \; University of Paris Saclay \qquad \qquad \qquad  INRIA  \qquad\qquad     }
\maketitle

\begin{abstract}
Differential privacy (DP)  and local differential privacy (LPD) are frameworks to protect sensitive information in data collections. They are both based on obfuscation. In DP the noise is added to the result of   queries on the dataset, whereas in LPD  the noise is added directly on the individual records, before being collected. The main advantage of LPD with respect to DP is that it does not need to assume a  trusted third party. The main disadvantage is that the trade-off between privacy and utility is usually worse than in DP, and typically to retrieve reasonably good statistics from the locally sanitized data it is necessary to have a huge collection of them. In this paper, we focus on the problem of estimating counting queries from collections of noisy answers, and we propose a variant of LDP based on the addition of geometric noise. Our main result is that the geometric noise  has a better statistical utility than other LPD mechanisms from the literature. 
\end{abstract}


\section{Introduction}

With the ever-increasing use of internet-connected devices, 
personal data are collected in larger and larger amounts, and then stored and manipulated  
for the most diverse purposes.  Undeniably, the big-data technology   provides enormous benefits to
industry, individuals and society. 
On the other hand, however, the collection and manipulation 
of personal data raises alarming privacy issues. Not surprisingly, therefore, 
the investigation of mechanisms to protect   privacy has become a very active field of research.


{Differential privacy} (DP) \cite{Dwork:06:TCC} and {local differential privacy} (LDP) \cite{Duchi:13:FOCS}
 represent the cutting-edge of research on privacy. 
DP aims at protecting the individuals' data while allowing to answer queries on the aggregate information, 
and it achieves this goal by adding controlled noise to the query outcome. 
LDP is a distributed variant   in which the data are sanitized at the user's end  before  being collected. 
One of the main reason of their success is that DP and LPD are \emph{compositional}, i.e., 
robust to  attacks based on combining the information from different sources. 
Furthermore LPD has the additional advantage that there is no need to assume that the entities collecting and storing data are  trusted, because they can only see, stock and analyze  the already sanitized data. 
 
LDP is having a considerable impact, especially now that large companies such as Apple and Google 
have  adopted it for collecting   their customers's data for statistical purposes. 


In this paper we consider the problem of \emph{statistical utility}, namely how precisely can we retrieve  the original distribution 
from the collection of noisy data.  Reconstruct the original distribution is important in order to make precise statistical analyses.  

The notion of $d$-privacy has been advocated in a recent work \cite{Alvim:18:CSF} as a variant of LDP  able to provide a good trade-off between privacy and statistical utility. 
In this paper, we consider a particular $d$-private mechanism: the geometric noise distribution. We explore its properties and we show that indeed, in terms of trade-off privacy-utility,  it compares favorably 
to the typical LPD mechanism, the $k$-Randomized-Responses ($k$RR)~\cite{Duchi:13:FOCS}.  

\section{Preliminaries}
In this section we recall some basic notions. We will consider only finite sets and discrete mechanisms. 
Given a    set ${\mathcal X}$, a probability distribution $p$ on ${\mathcal X}$ is a function $p:{\mathcal X}\rightarrow {\mathbb R}$
such that $\forall x\in {\mathcal X}\, p(x) \geq 0$ and $\sum_x p(x)=1$.  We   denote by 
${\mathit Distr}({\mathcal X})$ the set of  all possible   distributions on ${\mathcal X}$. We   
use $p_x$ to denote $p(x)$.
\subsection{Differential privacy}
Let $D, D'$   denote   collections of data (datasets),  ${\mathcal D}$ the set of all datasets of interest,  and let $\sim$   represent the \emph{adjacency relation} between datasets. Namely, 
$D\sim D'$ means that $D$ and $D'$ differ  only for the value of a single record.
Given a query $f: {\mathcal D}\rightarrow {\mathcal X}$, a mechanism  $\mathcal  K$ for $f$ is a probabilistic function which, for every $D$, 
gives a \emph{reported answer} $y\in {\mathcal Y}$ with a certain probability distribution that depends on the \emph{true answer} to the query.
Let $P[{\mathcal  K}(D)=y]$   denote the probability that $\mathcal  K$  applied to $D$ reports the answer $y$. 
We say that $\mathcal  K$
satisfies $\varepsilon$-DP, where $\varepsilon$ is a non-negative real number denoting the level of privacy, 
 if   for every pairs of adjacent datasets $D\sim D'$, and for every $y \in {\mathcal Y}$, we have:
 \begin{equation}\label{eqn:DP}
  P[{\mathcal  K}(D)=y]\leq e^{\varepsilon} \, P[{\mathcal  K}(D')=y].
  \end{equation}
 
 \subsection{Local differential privacy and  Randomized Responses}
In LDP the idea is that the mechanism obfuscates directly the value of the data rather than the answer to a query. 
In this setting,   let ${\mathcal X}$ denote the set of all possible values  for the data. A mechanism $\mathcal K$ is a probabilistic function which, for every $x\in{\mathcal X}$, 
returns a \emph{reported value} $y\in {\mathcal X}$ with a certain probability distribution that depends on the \emph{true value} $x$.
Let $P[{\mathcal  K}(x)=y]$   be the probability that $\mathcal  K$  applied to $x$ reports $y$. 
$\mathcal  K$ provides  $\varepsilon$-LPD if for all $x,x',y\in {\mathcal X}$  
we have:  
 \begin{equation}\label{eqn:LDP1}
 P[{\mathcal  K}(x)=y]\leq e^{\varepsilon} \, P[{\mathcal  K}(x')=y].
  \end{equation}
  
 A typical mechanism to implement LDP is the  Randomized Responses  ($k$RR), where $k$ represents the size of  ${\mathcal X}$. In its simplest variant it is defined as follows: 
  \begin{equation}
 P[k{\rm RR}(x)=y] =\left\{\begin{array}{ll}
\frac{e^{\varepsilon}}{k-1+e^{\varepsilon}} \;\; & y=x\\[2ex]
\frac{e^{\varepsilon}}{k-1+e^{\varepsilon}} & y\neq x 
\end{array}
\right.
\end{equation}

 \subsection{$d$-privacy}
In $d$-privacy, like in LDP,  mechanism obfuscates directly the value of the data. The main difference is that 
the domain $X$ is assumed to be a metric space, namely be endowed with a notion of distance 
$d:{\mathcal X}\times {\mathcal X}\rightarrow \mathbb{R}^{\geq 0}$, where $\mathbb{R}^{\geq 0}$ is the set of non-negative real numbers. 

A mechanism $\mathcal  K$ provides  $\varepsilon$-$d$-privacy if for every  $x,x',y\in {\mathcal X}$   
we have:
 \begin{equation}\label{eqn:LDP2}
 P[{\mathcal  K}(x)=y]\leq e^{\varepsilon\,d(x,x')} \, P[{\mathcal  K}(x')=y].
  \end{equation}

 \subsection{Generalized counting queries}
In DP,  a counting query is a function $f:{\mathcal D}\rightarrow [0,n]$ such that $f(D)$ gives the number of  records in $D$ that  satisfy a certain property. 
($[0,n]$  denotes the set of integers between $0$ and $n$.)
In this paper, we will adopt a more general notion of counting query, suitable for LPD. 
Namely, we assume that  $f:{\mathcal X}\rightarrow [0,n]$ associates a number $f(x)\in [0,n]$ to each element of $x\in X$. 
The idea is that  each $x\in X$ represents a certain person, and $f(x)$ could return, for example, the age (in years), or the number of children, or the monthly salary (in Euros), etc. 

A mechanism $\mathcal  K$ for $f$, in this context, associates to each value $i\in [0,n]$ a value  $j\in [0,n]$  chosen randomly 
according to a  probability distribution.
We   denote by $C_{ij}$ the probability that ${\mathcal K}(i)=j$. 
Note that $C_{ij}$ represent the conditional probability of $i$ given $j$, 
hence the values $C_{ij}$ form a stochastic matrix $C$ (where $C_{ij}$ is the element at the intersection of the $i$-th row and $j$-th column). 
From now on for notational simplicity we will 
use $C$ rather than $\mathcal K$. 

\subsection{Geometric mechanism}
In the following, for simplicity we  use $\alpha$ to indicate $e^{-\varepsilon}$, where ${\varepsilon}$ is the level of privacy. Note that $0< \alpha\leq 1$.
The geometric mechanism (for a counting query) is represented by an infinite matrix $C$ with rows indexed by $[0,n]$ and columns indexed by 
$\mathbb{Z}$ (the set of integers), and whose elements are given by:
\begin{equation}\label{eqn:geom}
C_{ij}=\frac{1-\alpha}{1+\alpha}\alpha^{|i-j|}
\end{equation}

In order to avoid dealing  with an infinite output domain, we consider the \emph{truncated} version of a  mechanism. The idea is that the probability mass of the negative element is \emph{remapped} in $0$, and the probability mass of the elements greater than $n$ is \emph{remapped} in $n$. 
The \emph{truncated geometric mechanism} will be denoted by $G$ and it is defined as:
\begin{equation}\label{eqn:tg}
G_{ij}=\left\{\begin{array}{ll}
\frac{1}{1+\alpha}\alpha^i &j=0\\[2ex]
\frac{1-\alpha}{1+\alpha}\alpha^{|i-j|} &0<j<n\\[2ex]
\frac{1}{1+\alpha}\alpha^{|i-n|} &j=n
\end{array}
\right.
\end{equation}
The truncated geometric is   $\varepsilon$-$d$-private:
\begin{proposition} \cite{Kacem:18:PLAS}
If ${\mathcal X}$ is the domain $[0,n]$ and $d$ is the difference between integers, then $G$ is a 
$d$-private mechanism on ${\mathcal X}$. 
\end{proposition}
The following is another important property of the truncated geometric:
\begin{proposition}\label{prop:inv}\cite{Kacem:18:PLAS}
The matrix $G$  is invertible.
\end{proposition}

\section{Reconstructing  the original distribution from a collection of noisy data}\label{sect4}
Assume that we have a collection of $N$  noisy data representing the  result of the independent application of the geometric mechanism 
to the data of a certain population. Each datum (as well as each noisy datum) is a number in $[0,n]$.
Let $\pi\in{\mathit Distr}([0,n])$ be the prior  distribution on the original data.
The set of original data is generated by a sequence of random variables $X_1,X_2,\ldots,X_N$ independent and identically  distributed (i.i.d.), according to $\pi$. 
%
%
To each of the  $X_1,X_2,\ldots,X_N$ we apply the geometric mechanism $G$, thus obtaining a sequence of random variables 
$Y_1,Y_2,\ldots,Y_N$. 
Let $q\in{\mathit Distr}([0,n])$ be the \emph{empirical}  distribution determined  by $Y_1,Y_2,\ldots,Y_N$. I.e., $q_j$ is obtained by counting the frequencies of the value  $j$ in $Y_1,Y_2,\ldots,Y_N$. Namely, $q_j=\nicefrac{|\{h\mid Y_h=i\}|}{N}$.

The task we consider here is how best to reconstruct the original distribution $\pi$ from $q$. 
To this purpose, we consider the following iterative procedure, which is inspired by the Bayes theorem. In the definition of this procedure,  $p$ represents an arbitrary probability distribution with full support. 
\begin{definition}\label{def:IBU}
Let $\{ p^{(k)}\}_k$ be the sequence  defined inductively as follows: 
\[
\begin{array}{rcl}
p^{(0)}&=&p\\[2ex]
p^{(k+1)}_i&=&\sum_j q_j\frac{p^{(k)}_i\alpha^{|i-j|}}{\sum_h p^{(k)}_h\alpha^{|h-j|}}
\end{array}
\]
\end{definition}
The interest of the above definition relies in the following result:
\begin{theorem}\cite{Kacem:18:PLAS}\label{theo:IBU}
Let $\{ p^{(k)}\}_k$ be the sequence of distributions constructed according to  Definition~\ref{def:IBU}. Then: 
\begin{enumerate}
\item
The sequence converges, i.e.,   $\lim_{k\rightarrow\infty}p^{(k)}$ exists.
\item 
$\lim_{k\rightarrow\infty}p^{(k)}$ is the \emph{Maximum Likelihood Estimator}  (MLE) of $\pi$ given $q$. 
\end{enumerate}
\end{theorem}
We will denote by $p^{*}$ the limit of the sequence $\{ p^{(k)}\}_k$, i.e.,  $p^*\stackrel{\rm def}{=}\lim_{k\rightarrow\infty}p^{(k)}$. 
Theorem~\ref{theo:IBU}(2) means that  for all possible distributions $p'$, the probability that the distribution induced from the 
noisy data (sanitized with $G$) is  $q$ when the prior is $p^*$ is higher than or equal to the same probability when the 
prior is $p'$. 

Furthermore, $p^*$ can be characterized  using $G$. For  a distribution   $p$ and a matrix $C$, let $p C$ be the product of  $p$ and $C$.  
Namely, $(pC)_j= \sum_i p_iC_{ij}$. 
\begin{proposition}\cite{Kacem:18:PLAS}
If $r = q\,G^{-1}$ is a probability distribution, then $p^* = r$.
\end{proposition}

\section{Comparison between the Geometric and Randomized Response mechanisms}
In this section we compare the truncated geometric and the $k$RR mechanisms from the point of view of the trade-off between privacy and 
statistical utility. 

In order to make a fair comparison, we first need to calibrate the privacy parameters of these mechanisms  so that they represent the same level of privacy. Indeed, although both are expressed in terms of a parameter $\varepsilon$, they do not have the same meaning: the first satisfies ${\varepsilon}$-$d$-privacy, while the second satisfies ${\varepsilon}$-LPD.

To demonstrate, consider the $k$RR mechanism with parameter ${\varepsilon}=\ln 2$ operating over integer-valued input and output domains with range $[0, 100]$.
The privacy guarantee provided by this mechanism is given by the upper bound $\varepsilon^{\ln 2} = 2$, representing the maximum likelihood ratio between any possible reported value and the true value. This upper bound is realised for every pair of different values in the input and output domains.
By comparison, the truncated geometric mechanism with the same $\varepsilon={\ln 2}$ would provide such an upper bound $2$ only for values immediately adjacent to the true one. For values further away, the bound is smaller (making more distance values less likely).
If we want to provide the same upper bound $2$ on the entire domain, then we would have to set $\varepsilon$ to a value $100$ times smaller, namely $\nicefrac{\ln 2}{100}$, which  
would result in a very flat curve, making the true value almost indistinguishable from a large part of the other values. 

\begin{figure}[t]
        \center{\includegraphics[width=\textwidth]  {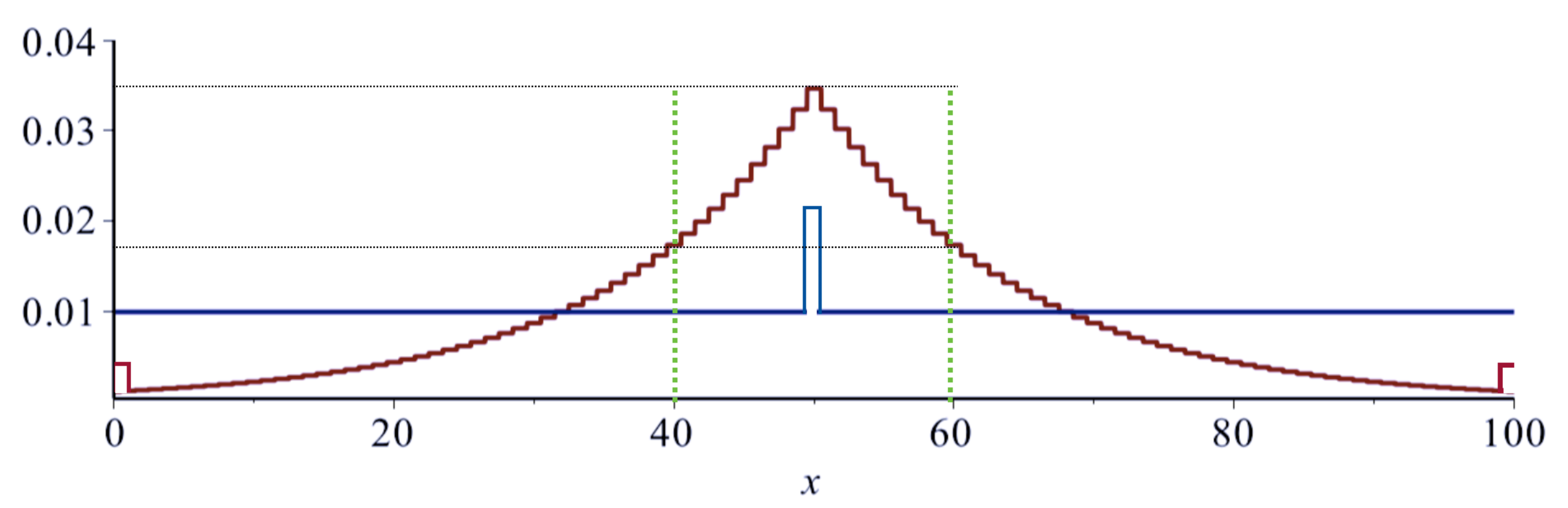}}
        \caption{\label{fig:LDP} The distribution generated by the  $k$RR  and the truncated geometric  mechanisms applied to $x=50$. The values of the privacy parameters $\varepsilon$'s are $\ln 2$ and $\nicefrac{\ln 2}{10}$, respectively.}
      \end{figure}

However, we argue that it is not necessary to inject so much noise, as this destroys the utility-by-design of the geometric mechanism. As a compromise we will require the upper bound $2$ on a restricted subset of elements, for instance those in a radius  $10$ from the true value. This can be achieved by setting $\varepsilon$ to $\nicefrac{\ln 2}{10}$. Figure~\ref{fig:LDP} illustrates the situation.
 
As for statistical utility, intuitively it should account for how well we can approximate   statistics  on the original data by using only the collected noisy data. This can be formalized in terms of the distance between the original distribution and the most likely one given the noisy data, which can be estimated by applying the IBU (Definition~\ref{def:IBU}). 
As for the notion of distance, we  propose to use the Kantorovich metric (based on the standard distance between natural numbers as the ground distance). 
As argued in  \cite{Alvim:18:CSF}, in fact, this metric is related to a large class of statistical functions. 
We recall the definition of the Kantorovich distance: 
\begin{definition}
Let $({\cal X},d)$ be a metric space and let $\mu,\mu'\in {\mathit Distr}({\mathcal X})$. The Kantorovich distance based on $d$ between $\mu$ and $\mu'$ is defined as follows: 
\[
K_d(\mu,mu') = \max_{g\in\mathscr{G}} \mid \sum_{x\in {\cal X}}g(x)\mu(x) - \sum_{x\in {\cal X}}g(x)\mu'(x)
\]
where $\mathscr{G}$ is the set of the Lipshitz functions on ${\cal X}$, namely $g\in \mathscr{G}$ if and only if $\forall x, x' \in {\cal X}  \mid f(x) - f(x') \mid \leq d(x,x') $. 
\end{definition}



\subsection{Experimental Results}

We now present the results of experiments designed to assess the statistical utility of each of these mechanisms using the IBU method outlined in Section~\ref{sect4}.

As above, we assume integer-valued inputs and outputs in the range $[0, 100]$. We constructed
two different mechanisms to output noisy values: a truncated geometric mechanism parametrised by $\varepsilon = \nicefrac{ln 2}{10}$ and a $k$RR mechanism parametrised by $\varepsilon = {ln 2}$. 

We ran our experiments on 2 sets of data. The first set consisted of samples of size 1000, 10000, 50000 and 100000 drawn from a binomial distribution. The second set consisted of the same sample sizes drawn from a ``4-point'' distribution (i.e. a random distribution over 4 `points' in the output range). For each of the 8 samples we conducted 20 experiments using the following method:
\begin{enumerate}
\item Obfuscate the sample using each of the (geometric and $k$RR) mechanisms to produce 2 obfuscated sets.
\item Convert each set into an empirical distribution over outputs using the frequency counts of elements in each set.
\item Run IBU for 5000 iterations over each empirical distribution to compute the maximum likelihood estimate (MLE) for the true distribution.
\item Compare the Kantorovich distance between the MLE and the true distribution as an estimate of the error caused by the obfuscation.
\end{enumerate}

\begin{figure}
	\centering	
	\mbox{
	\subfigure{
		\includegraphics[width=0.5\linewidth]{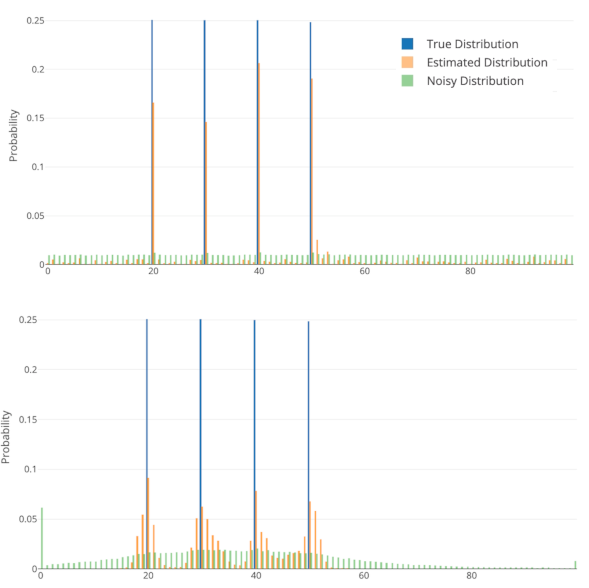}}
	\subfigure{
		\includegraphics[width=0.5\linewidth]{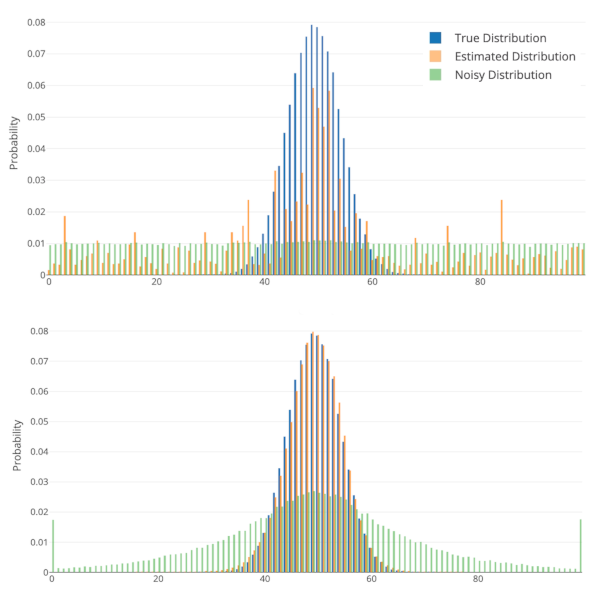}}
	}
	\caption{\label{fig:ibu-approx}IBU reconstruction of MLE (orange) distributions from noisy (green) distributions based on 100k samples drawn from `4-point' (left) and binomial (right) distributions. The blue graphs indicate the true distribution. The top distributions were obfuscated by kRR, and the bottom by the geometric mechanism. Reconstruction for the kRR is much better for the point distribution, but the opposite is true for the geometric mechanism.}
\end{figure}

In Figure~\ref{fig:ibu-approx} we present some sample runs of IBU for each mechanism and distribution. Interestingly, the reconstructed distribution for kRR is much better for the `4-point' sample than for the binomial sample. Conversely, the reconstructed distribution for the geometric mechanism is much closer to the binomial sample.

\begin{figure}[t]
        \center{\includegraphics[width=\textwidth] {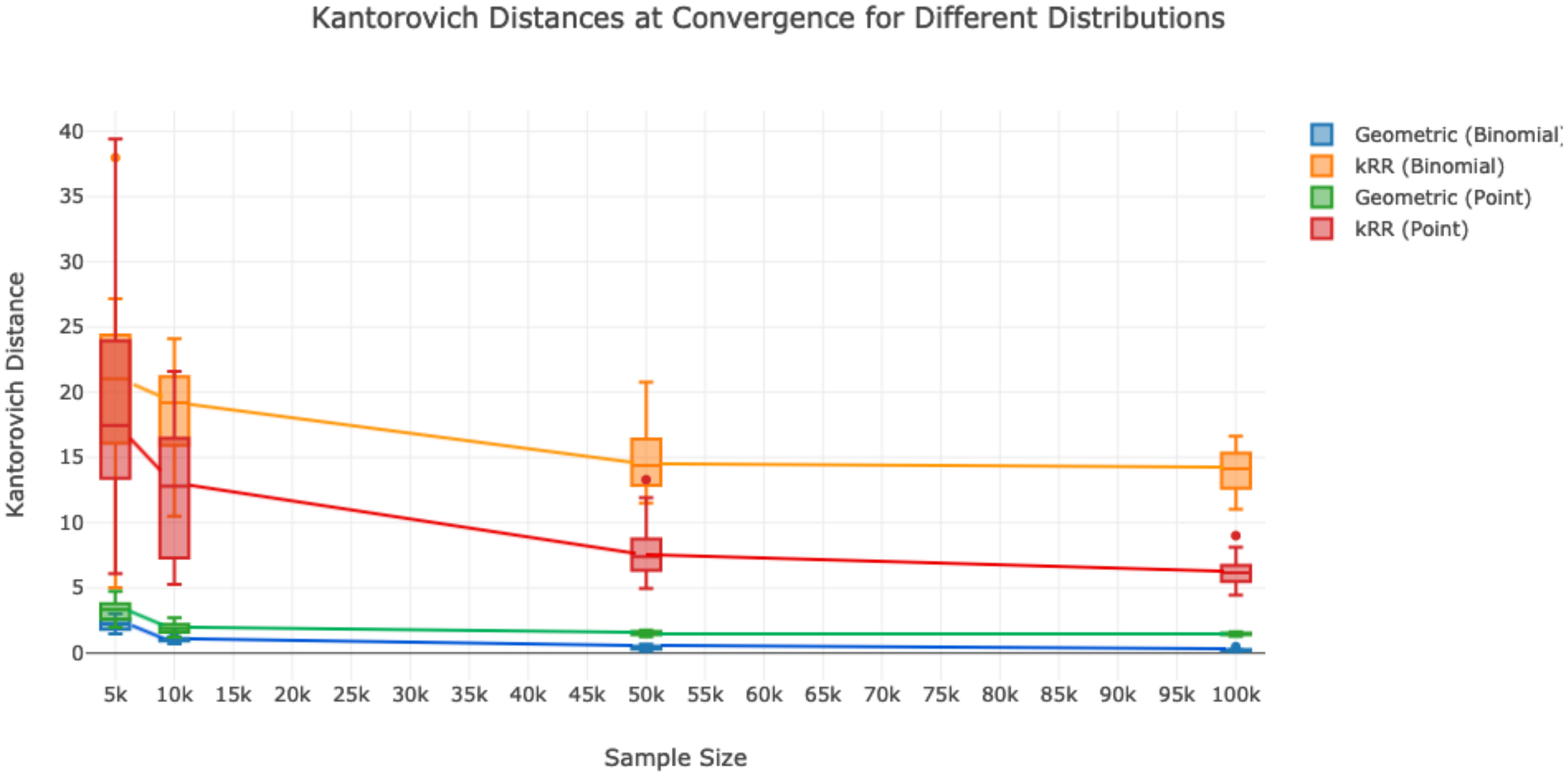}}
        \caption{\label{fig:ibu-results} Kantorovich distances between true and estimated distributions at IBU convergence for the geometric and kRR mechanisms. Distances were computed over 20 experiments for each of the 4 sample sizes indicated. This shows the distributions produced by the geometric mechanism are much closer to the true distribution than for the kRR.}
\end{figure}

However, the computed Kantorovich distances at the 5000 iteration point for each run tell a different story. These results are shown in Figure~\ref{fig:ibu-results}. We computed the Kantorovich distance between the estimated distribution and the true distribution, providing an approximation of the distance between the true distribution and the distribution resulting from obfuscation. We can see that the average Kantorovich distances for the geometric mechanism are significantly lower (up to 5 times) than the corresponding distances for the kRR mechanism. We conjecture that this is because the errors caused by kRR are randomly distributed over the entire output space, which directly affects the Kantorovich distance since it depends on the ground distance between points. This means that for statistical applications in which the ground distance is important, the geometric mechanism is still preferred to the kRR mechanism.

\begin{figure}[h!]
        \center{\includegraphics[width=0.8\textwidth] {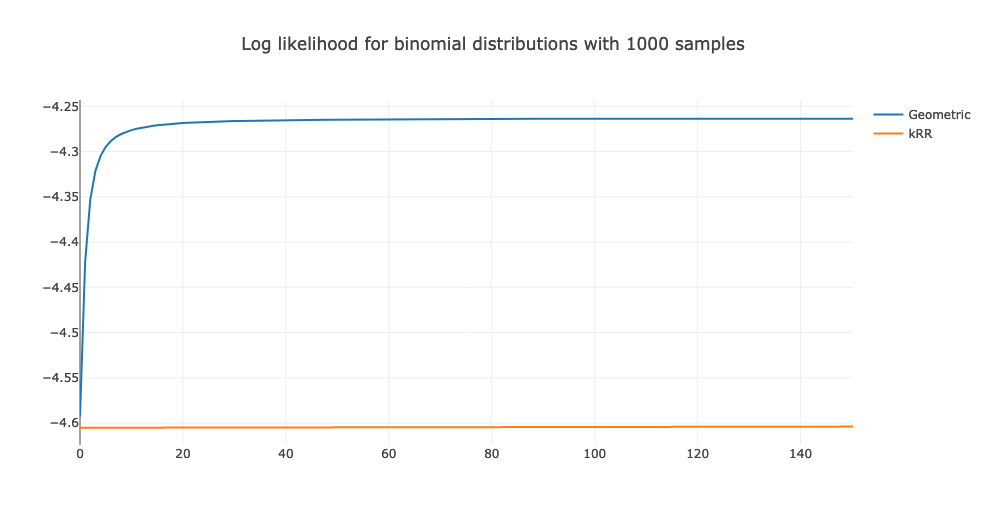}}
        \caption{\label{fig:convergence} Log likelihood function against number of iterations for the geometric and kRR mechanisms. This graph shows
        how fast each output distribution converges to the MLE for one particular (representative) run of the IBU. We observe that the geometric mechanism converges quickly whereas convergence for the kRR is almost flat.}
\end{figure}

Another interesting observation we make is in the convergence rates for the IBU method when applied to the different distributions. This is graphed in Figure~\ref{fig:convergence}. For each iteration of IBU we computed the `log likelihood' function
\[
            L(\Theta) = \sum_y q_y \log(\Theta \cdot M_y)
\]
where $\Theta$ is the current estimated distribution, $q_y$ is the empirical distribution and $M$ is the mechanism represented as a channel matrix. ~\footnote{The notation $\Theta \cdot M_y$ indicates the dot product of $\Theta$ with the $y$th column of $M$.} The log likelihood function indicates how close the current estimate is to the true MLE. The results for one particular run are shown in Figure~\ref{fig:convergence}. We can see that the geometric mechanism converges to a close approximation of the MLE within 10 iterations, whereas the convergence for kRR is linear and almost flat. This may also explain the better performance of the kRR output on the `4-point' sample, since there were far fewer `skyscrapers' in the original distribution to estimate. The shape of the geometric mechanism seemed to favour the more `natural' shape of the binomial distribution sample.

\section{Conclusion}
In this paper, we have investigated the properties of the truncated geometric mechanism in relation to the  reconstruction from noisy data of the original distribution on the real data. 
We have provided an iterative algorithm to approximate the original distribution, and we have given a characterization of 
the fixed point in terms of the inverse of the matrix. Finally, we have compared the trade-off between privacy and utility of the the truncated geometric mechanism and of the kRRs, obtaining favorable results. 

\balance

%
%
%

 \subsection*{Acknowledgements}                            
The work of Catuscia Palamidessi has been partially supported by  the
ANR  project  {REPAS}. 

\bibliography{biblio_cat}

%

\end{document}